\documentclass[%
 reprint,
superscriptaddress,
 amsmath,amssymb,
 aps,
 twocolumn,
]{revtex4}

\usepackage{graphicx}
\usepackage{dcolumn}
\usepackage{bm}

\usepackage{float}
\usepackage[dvipsnames]{xcolor}
\usepackage{soul}
\usepackage{appendix}

\begin{document}


\title{Surface tensiometry of phase separated protein and polymer droplets by the sessile drop method}

\author{Mahdiye Ijavi}
\affiliation{Department of Materials, ETH Z\"{u}rich}%
\author{Robert W. Style}%
\affiliation{Department of Materials, ETH Z\"{u}rich}%
\author{Leonidas Emmanouilidis}%
\affiliation{Institute of Biochemistry, ETH Z\"{u}rich}%
\author{Anil Kumar}%
\affiliation{Laboratory of Biomolecular Research, Paul Scherrer Institute}%
\author{Sandro M. Meier}%
\affiliation{Institute of Biochemistry, ETH Z\"{u}rich}%
\affiliation{Laboratory of Biomolecular Research, Paul Scherrer Institute}%
\author{Alexandre L. Torzynski}%
\affiliation{Department of Materials, ETH Z\"{u}rich}%
\author{Fr\'{e}d\'{e}ric H. T. Allain}%
\affiliation{Institute of Biochemistry, ETH Z\"{u}rich}%
\author{Yves Barral}%
\affiliation{Institute of Biochemistry, ETH Z\"{u}rich}%
\author{Michel O. Steinmetz}%
\affiliation{Laboratory of Biomolecular Research, Paul Scherrer Institute}%
\author{Eric R. Dufresne}%
 \email{eric.dufresne@mat.ethz.ch}
\affiliation{Department of Materials, ETH Z\"{u}rich}

\date{\today}

\begin{abstract}
Phase separated macromolecules play essential roles in many biological and synthetic systems. 
Physical characterization of these systems can be challenging because of limited sample volumes, particularly for phase-separated proteins.
Here, we demonstrate that a classic method for measuring the surface tension of liquid droplets, based on the analysis of the shape of a sessile droplet, can be effectively scaled down   for this application.
The connection between droplet shape and surface tension relies on the density difference between the droplet and its surroundings. 
This can be determined with small sample volumes in the same setup by measuring the droplet sedimentation velocity.
An interactive MATLAB script for extracting the capillary length from a droplet image is included in the ESI.
\end{abstract}

\maketitle

\section{\label{sec:Intro}Introduction}

Solutions of macromolecules  have long been known to phase separate into liquid domains that share the same solvent but differ in their composition \cite{ostwald1927, de1932, overbeek1957}.  
An essential feature of these droplets is their extremely low surface tension, $\gamma$.
Classically, this can be determined by the spinning drop method \cite{liu} where a relatively low density droplet is placed in a horizontal capillary tube that is spun very rapidly along its axis of symmetry.
By quantifying the shape of the droplet as a function of the rotation rate, the surface tension can be determined with stunning accuracy, with reports of surface tensions as low as $10^{-6}~\mathrm{ N/m}$. 

Liquid-liquid phase separation of macromolecular solutions has enjoyed intense interest in recent years thanks to the discovery that  many membrane-less compartments  within eukaryotic cells appear to be liquid droplets of phase-separated macromolecules \cite{hymn1, web}.
Examples include P granules \cite{clif2}, P bodies \cite{ked1}, Cajal bodies \cite{stz} and nucleoli \cite{clif3}. 
Membraneless organelles have complex compositions, including a large number of distinct  RNAs and proteins.
The bulk of the proteins found in these droplets are intrinsically-disordered and lack globular structure \cite{wright,dunker,dyson,tomp}. 
These proteins readily phase separate \emph{in vitro} to form a dilute `supernatant' phase and a protein-rich phase,
which can take the form of liquid droplets, hydrogels, fibrils or aggregates \cite{vek}.
The tendency  of intrinsically disordered proteins to phase separate is thought to  to underlie the formation of membraneless compartments in the cell \cite{web2,ked2,tomp2}.
Precise physical characterisation of droplets \emph{in vitro} can provide insights into the molecular scale interactions of their components, and potentially shed light on their physiology \emph{in vivo}.

The essential challenge in characterizing the properties of these droplets is their size.
Typically no more than a few microns across \emph{in vivo}, larger droplets can be formed \emph{in vitro}.
However, the production and purification of protein is labor intensive and expensive, and total volumes of the protein-rich phase are typically limited to the $\mathrm{\mu L}$ scale.
Such samples are too small for conventional mechanical measurements.
For example, quantification of bulk mechancial properties through conventional rheometry requires at least $100~\mathrm{\mu L}$ of material.
In the last twenty years, a number of new approaches have been developed to scale-down rheological measurements  \cite{furst2017}.
Among these, particle tracking microrheology is well-suited to measure the linear rheology of phase separated droplets \cite{Taylor,Murakami}.

Similarly efficient and effective approaches have not yet been established to characterize the surface mechanical properties of droplets.
Classic spinning drop measurements can work for droplets on the $\mathrm{\mu L}$ scale, but only if they are less dense than their surroundings. 
In principle, the spinning droplet method could be applied to a small droplet of the protein-depleted phase within a larger volume of the protein-rich phase, but that would require  volumes of the protein-rich phase around $1000~\mathrm{\mu L}$.
A number of methods have been introduced in recent years to estimate the surface tension of micron-scale phase separated droplets.
The most popular method is based on measuring the time it takes for two droplets to fuse \cite{shanacliff}.
In principle, this method can be very accurate if the rheology of the droplets is known.
While fusion times are easy to calculate from first principles in bulk, they can be strongly modified when droplets are in contact with a surface, as contact lines can dramatically slow down the movement of any  droplet \cite{hernandez2012symmetric}.
Imaging of fusion events in the absence of contact lines is very challenging. 
At droplet volume fractions low enough to clearly resolve the dynamics, fusion events  are rare.
Optical tweezers can be used to initiate fusion, but care has to be taken to avoid contributions of optical forces to the fusion time as well as photo- and thermal damage to the droplets \cite{alfonso}.
Alternatively,  measurements  of the force required to statically deform a droplet can be used to accurately determine the surface tension using  optical tweezers \cite{Jawerth} or atomic force microsopy \cite{zitzler2002capillary,anachkov2016} .

Alternatively, the shape of droplets of an appropriate size contains sufficient information to accurately determine their surface tension.  
For conventional simple liquids, made of small molecules, this is done through the analysis of photographs of millimeter scale pendant or sessile droplets \cite{saad2016,Atefi}.
Surface tension is most easily inferred from the shape of a droplet when its radius is larger than the capillary length, $L_c=\sqrt{\gamma/\Delta \rho g}$.
Here, $\Delta \rho$ is the density difference between a droplet and its surroundings and $g$ is the acceleration due to gravity.
The tiny values of surface tension for phase-separated droplets of macromolecules are somewhat mitigated by a  simultaneous reduction of the density difference.  
All together, this leads to reported values of the capillary length varying from $\mathcal{O}(10~\mathrm{\mu m})$ to $\mathcal{O}(100~\mathrm{\mu m})$ \cite{liu}. 
This corresponds to droplets with volumes of  the order $\mathcal{O}(\mathrm{nL})$. 
By analyzing the shape of fused fluorescently tagged nucleoli in giant \emph{Xenopus} oocytes, Feric \emph{et al} determined their capillary length to be of order $\mathcal{O}(10~\mathrm{\mu m})$ \cite{Feric}
One challenge of this approach is that the density difference between the droplet and surrounding fluid needs to be known accurately.
This can be inferred from sedimentation speeds of microscopic droplets when the viscosities of the droplet and surroundings are known \cite{Feric2}.

Here, we apply established methods of sessile droplet shape analysis to determine the surface tension of phase-separated macromolecular droplets with unknown density.
Using a standard camera equipped with a telecentric lens, we quantify phase separated liquid droplets, made of polymers or proteins, with surface tensions ranging from 7 to 90 $\mathrm{\mu N/m}$.
We determine the density difference in the same setup by measuring the droplets' sedimentation velocity.  
Typical values of the density difference range from 30 to 150 $\mathrm{kg/m^3}$.
This approach is attractive for the \emph{in vitro} measurement of phase-separated proteins because it uses small sample volumes and the proteins do not need to be fluorescently tagged.

\section{Droplet Preparation and Imaging}

We determine the surface tension of phase-separated macromolecular liquids from the shape of sessile droplets (Sec. \ref{sec:sessile}) and the speed of sedimenting droplets (Sec. \ref{sec:sediment}).
For a meaningful surface tension measurement, these droplets must be immersed in a co-existing liquid phase.
Typically, the denser macromolecule enriched phase forms the droplet, and the macromolecule-poor phase forms the continuous phase.
These coexisting phases are readily separated by centrifugation.

\begin{figure}[t]
\centering
\includegraphics[width = .9 \columnwidth]{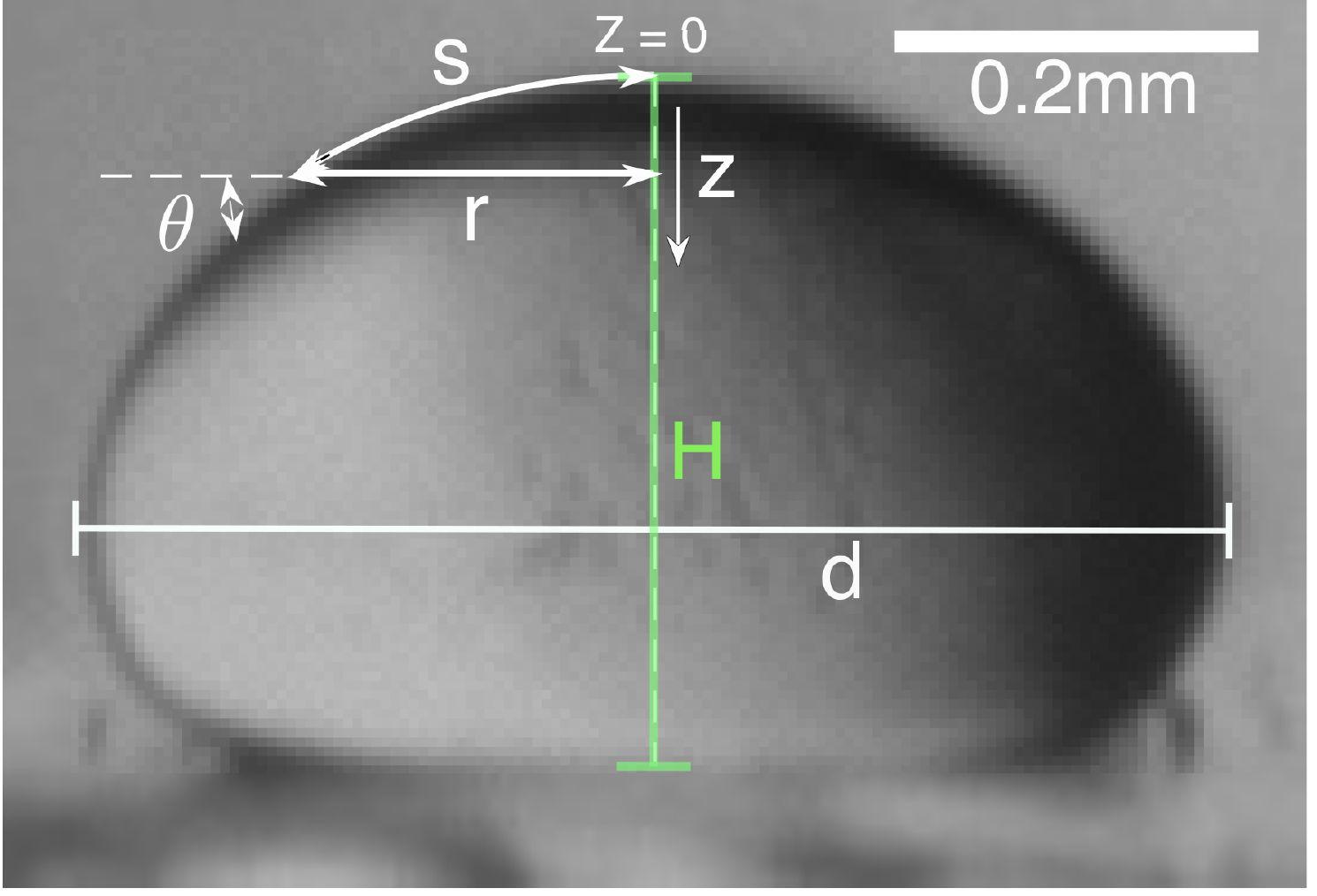}
\caption{\label{fig:sitdrop} Image of a sessile dextran-rich droplet in a continuous PEG-rich phase  on a PEG-ylated glass surface. $d$ is maximum width of the droplet and $H$ is the height of the droplet from the contact surface.}
\end{figure}

To image droplets, we
use a typical optical setup for sessile droplet analysis \cite{sefiane,drelich}, described in the Materials and Methods.
Droplets are  illuminated by an extended light source and imaged with a telecentric lens. 
To limit artifacts due to refraction, the droplet and continuous phases are held in a rectangular cuvette.
A typical image of dextran-rich droplet in a PEG-rich continuous phase is shown in Fig. \ref{fig:sitdrop}.
A series of droplets of different sizes of the same material are shown in the inset to Fig. \ref{fig:sitdropbonhd}.
The droplet shape, $r_\mathrm{exp}(z)$, is manually extracted from the acquired images in MATLAB, using the script included in the ESI.

Large droplets, appropriate for surface tension measurements, described in Section \ref{sec:sessile},  can be  formed by direct pipetting of the droplet phase into its coexisting continuous phase.
Alternatively,  a  droplet dispersion can be pipetted into the continuous phase.  The latter approach allows for a wider range of droplet sizes, including smaller droplets appropriate for  sedimentation velocity measurements of Section \ref{sec:sediment}.

To accurately determine the 3D shape of the droplet from such 2D images,  droplets must be axisymmetric.
As we will see in Section \ref{sec:sessile}, the accuracy of the surface tension measurement is best for droplets with a large contact angle.
Phase separated droplets of proteins often interact strongly with standard glass and plastic surfaces, spreading to very low contact angles and adopting non-axisymmetric shapes due to pinning.
Therefore, the bottom of the cuvette needs to be very clean and functionalized to minimize spreading.
As described in the Materials and Methods, a PEG-silane treated coverslip placed at the bottom of the cuvette is sufficient for the droplets studied here.

\section{Droplet Shape Analysis}
\label{sec:sessile}

The equilibrium shape of a liquid droplet is determined by a balance of surface tension and hydrostatic pressure.
For a droplet with cylindrical symmetry, this is captured by the axisymmetric Laplace equation \cite{del1997axisymmetric}:
\begin{equation}
    \label{eq:pg}
      \Delta \rho gz = P^* - \gamma \left({d\theta \over ds}+{\sin \theta \over r}\right),
\end{equation}
The coordinate system, including $s,~r,~z,~\theta$, is defined in Fig. \ref{fig:sitdrop}.
The left-hand side of the equation gives the contribution from gravity, which increases linearly from the top of the droplet.
The right-hand side of the equation gives the contribution of surface tension to the droplet pressure, which is determined by the local curvature.
The constant  $P^*=2\gamma/R_0$ is the pressure at the top of the droplet, where $R_0$ is the local radius of curvature.

To simplify the equation, we  scale $r,~s$ and $z$ by the capillary length  and rearrange to obtain
\begin{equation}
    \label{eq:num}
    {d\theta \over d\bar{s}} = 2\beta-\bar{z} - {\sin \theta \over \bar{r}},
\end{equation}
where  over-bars indicate non-dimensional lengths, and $\beta=L_c/R_0$ (the inverse square root of the standard Bond number).
This is now a one-parameter shape equation that we solve numerically with the geometric requirements:
\begin{equation}
    \frac{d\bar{r}}{d\bar{s}}=\cos\theta,\quad \frac{d\bar{z}}{d\bar{s}}=-\sin\theta,\quad {\sin \theta \over \bar{r}} (r\rightarrow 0) = {d\theta \over d\bar{s}}(r=0).
    \label{eqn:bcs}
\end{equation}
The last equation holds at the top of the drop, and represents the fact that the two principle surface curvatures are equal at this point \cite{del1997axisymmetric}.

\begin{figure}[t]
\includegraphics[width = .95 \columnwidth]{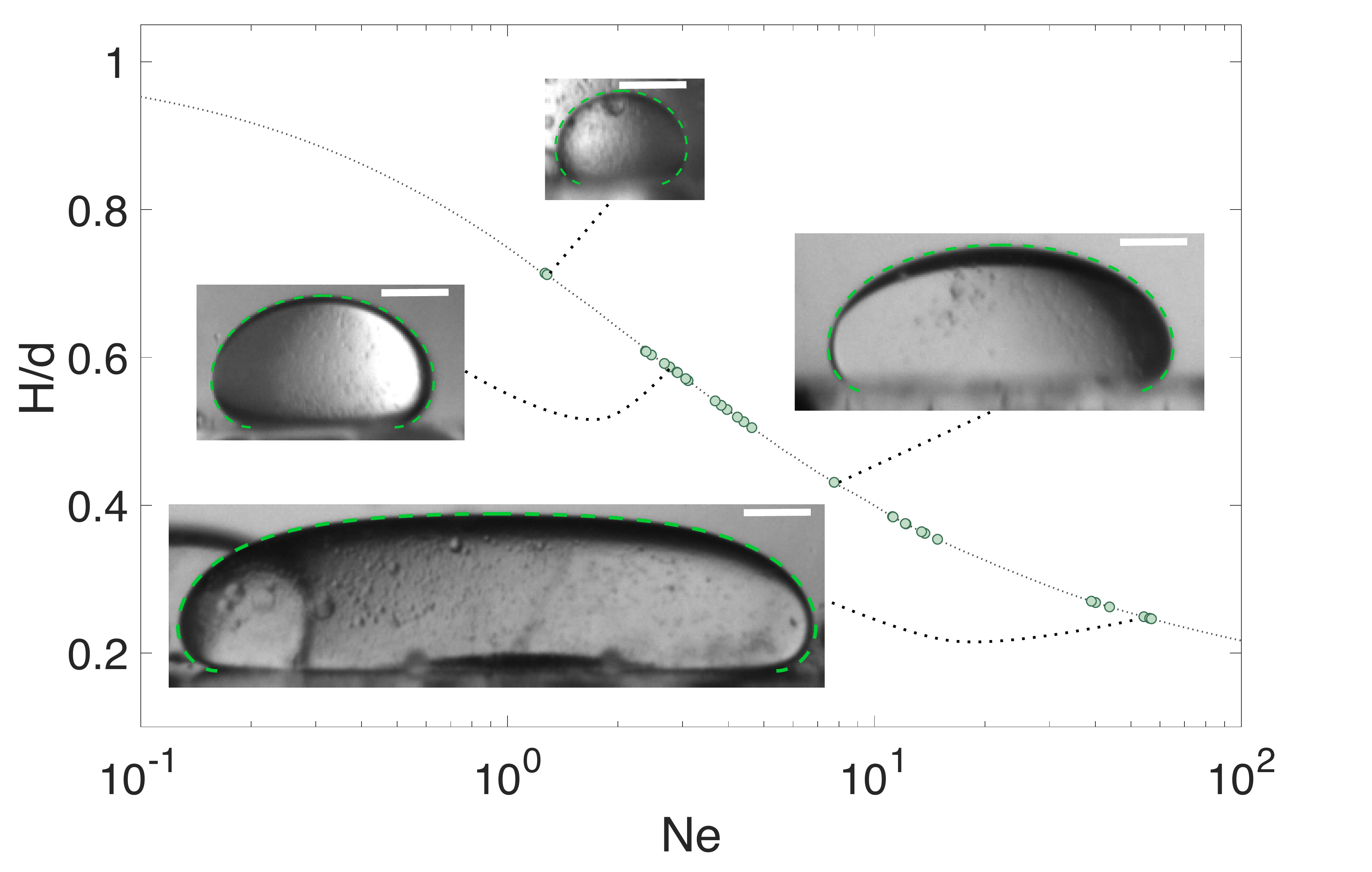}
\caption{\label{fig:sitdropbonhd} \emph{Sessile droplet shape analysis} Images of sessile dextran-rich droplets in a PEG-rich continuous phase. The scale bars are 200 $\mu m$.
Dashed-green lines on the droplets shown best fit curves to the droplet shape.  These images are superimposed over a scatter plot of the Neumann number and aspect ratio, $H/d$.  The dashed-grey curve is Neumann number in the perfect case where the sessile droplet makes 180 contact angle to the surface. }
\end{figure}

To extract the material properties of the droplet, we find the values of $\beta$ and $L_c$ that minimise the difference between the numerical solution, and the non-dimensionalised shape of the droplet, $r_\mathrm{exp}/L_c$.
Examples of the fits to droplet shape superimposed on raw droplet images are found in the insets to Fig. \ref{fig:sitdropbonhd}. 
An interactive MATLAB script for fitting droplet shape is found in the ESI.

Previous work has shown that  sessile droplet measurements of the surface tension are only accurate when the Neumann number, $\mathrm{Ne}=R_0H/L_c^2=H/(L_c\beta)>0.3$, where $H$ is the height of the droplet from the substrate \cite{yang2017accuracy}.
For smaller Ne, droplets are effectively spherical, and it is not possible to extract a meaningful value of $\gamma$.
Practically speaking, Ne can be increased by increasing droplet size, or by increasing the contact angle of the droplet.
A scatter plot of the Neumann number and droplet aspect ratio for a series of dextran-rich droplets in  a PEG-rich continous phase are shown in Fig. \ref{fig:sitdropbonhd}.

\section{Sedimentation Velocity}
\label{sec:sediment}

The sessile droplet measurement gives us the capillary length, $L_c=\sqrt{\gamma/\Delta\rho g}$.
To extract the surface tension, we need to know the density difference between the droplet and continuous phases. 
For relatively inexpensive molecules like PEG and dextran, milliliter scale samples of each phase can easily be produced, enabling standard density measurements of each phase \cite{Atefi}. 
For proteins, we require a technique that can be performed with sub-microlitre scale droplets.

\begin{figure}[t]
\centering
    \includegraphics[width = .95 \columnwidth]{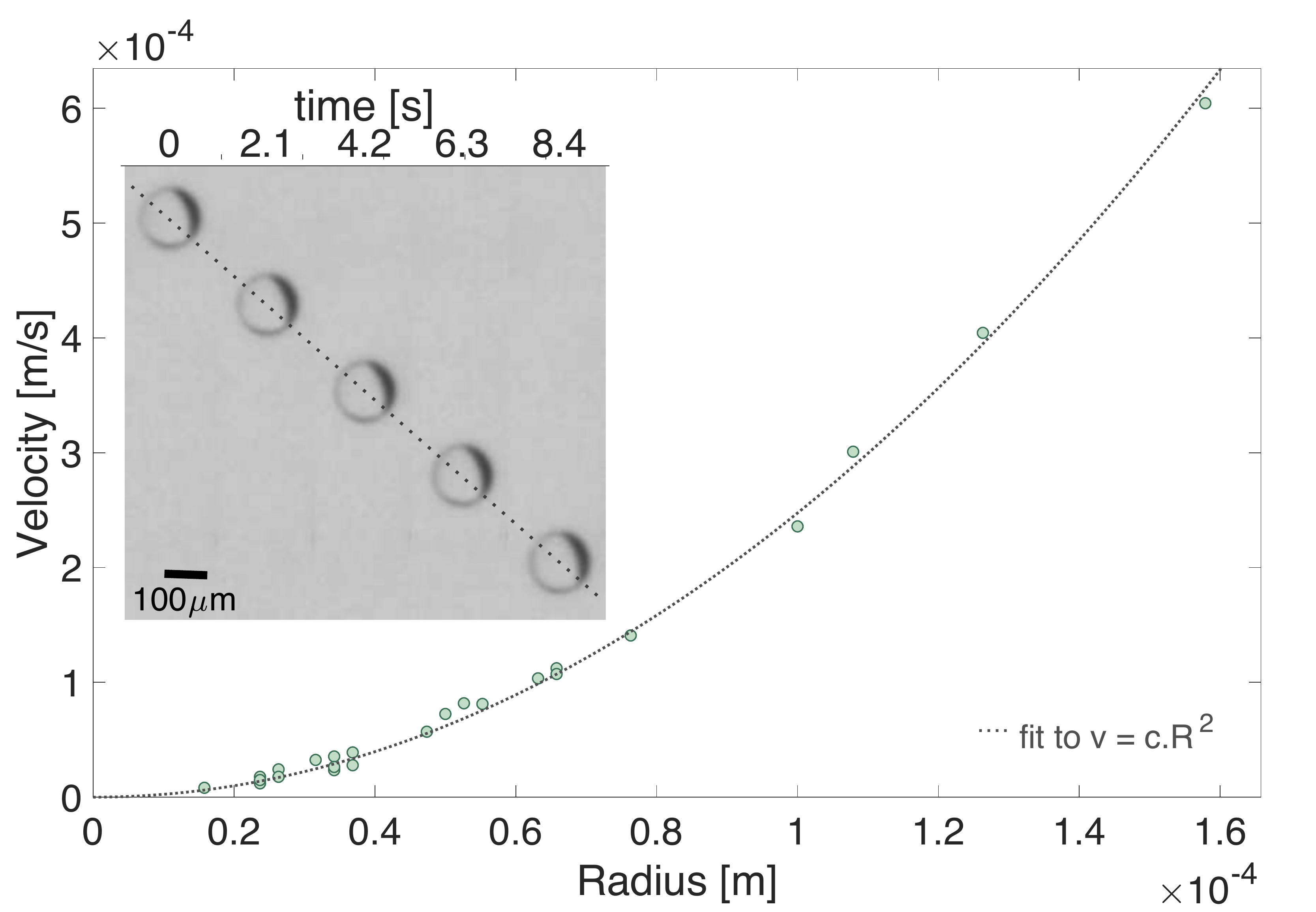}
    \caption{ \emph{Sedimenting droplets.}  Inset shows a timeseries of a dextran-rich droplet sedimenting in a PEG-rich continuous phase.  Each image is separated by 2.1 sec. The dashed line through the droplet centers shows that the sedimentation speed is linear. The main plot shows sedimentation speeds of 24 such droplets plotted against droplet radius.}
    \label{fig:falldrop}
\end{figure}

The density difference between the two phases can easily be determined by measuring the sedimentation speed of small droplets in the same setup used for the sessile droplet shape experiments. 
Example data for the PEG-dextran system is shown in Fig. \ref{fig:falldrop}.
Images of a sedimenting dextran-rich droplet, acquired at equally spaced time intervals are shown in the inset.
As shown by the straight line through their centers, these droplets move at a steady sedimentation velocity.
The sedimentation speed depends on the droplet size, and is plotted over a factor of eight in droplet radius in Fig. \ref{fig:falldrop}.

The sedimentation velocity, $v$, is determined by a balance of the buoyant force and viscous drag.
For a spherical droplet of radius $R$, the buoyant force is  $4 \pi \Delta \rho g R^3 / 3$. 
When the droplet viscosity, $\eta_d$, is much larger than the viscosity of the continuous phase, $\eta_c$,  the viscous drag is given by the familiar Stokes form, $6 \pi \eta_c R v$.   
Balancing the forces and solving for the density difference, we find
\begin{equation}
    \label{eq:stokes}
  \Delta \rho = \frac{9}{2} \frac{v}{R^2} \frac{\eta_c}{g}.
\end{equation}
If the continuous viscosity is comparable or larger than the droplet viscosity, the previous equation is corrected by a factor that depends on the viscosity ratio,
 \cite{hadamard1911mouvement}:
\begin{equation}
    \label{eq:modStokes}
    \Delta \rho = \frac{9}{2} \frac{v}{R^2} \frac{\eta_c}{g} \left(\frac{1 + (2/3) (\eta_c/\eta_d)}{1+(\eta_c/\eta_d)}\right).
\end{equation}
Note, however, that this correction is never more that a 33\% change in density.

Since macromolecule-rich fluids tend to be much more viscous than simple liquids,  Eq. \ref{eq:stokes} is usually quite accuruate for protein and coacervate droplets in their typical buffers.
However, for systems with signficant macromolecule concentrations in both phases Eq. \ref{eq:modStokes} may be necessary.
In such cases, independent measurements of the viscosities of both phases are suggested.
For the PEG-dextran example of Fig. \ref{fig:falldrop}, $\eta_c/\eta_d=0.05$, so that Eq. \ref{eq:stokes} is accurate to $<2\%$.

For fixed compositions of the droplet and continuous phases, the sedimentation velocity should scale like the square of the droplet radius.
This can be verified  by rearranging Eqs. \ref{eq:stokes} or \ref{eq:modStokes}.
The dashed-line in Fig. \ref{fig:falldrop} shows a good fit of the sedimentation speed to the $R^2$ form over this range of droplet radii.
Systematic deviations from this size-dependence can indicate that the droplets are not an appropriate size.

On the one hand, droplets need to be significantly bigger than  the resolution of the imaging system.
On the other hand, the droplets have to be small enough that they maintain a spherical shape while falling.
To remain spherical, droplets require the capillary number, Ca $=v\eta_c/\gamma\ll 1$ \cite{rallison1984deformation,stone1994}.
Inserting this into Eq. (\ref{eq:stokes}), this requirement becomes that $R \ll L_c$.
This criterion contradicts the requirement for accurate sessile droplet measurements, $R \gtrsim L_c$
Therefore, one should not use the same droplets for sedimentation velocity and surface tension measurements.  
For that reason, we prefer to pipette a premixed emulsion into the continuous phase.
However, when multiple droplets fall near each other, hydrodynamic interactions can dramatically effect their sedimentation velocity \cite{davis1985sedimentation,segre1997long}.
Therefore, sedimentation measurements should be limited to extremely dilute, well-separated droplets.  

The sedimentation of large droplets can also be impacted by inertial effects, quantified by the Reynolds number, Re $=\rho_c v R/\eta_c$.
 Equations \ref{eq:stokes} and \ref{eq:modStokes} are only valid when $\mathrm{
 Re}\ll 1$.
For the largest expected density differences (around $500~\mathrm{kg/m^3}$) in the least viscous continuous phase (an aqueous continuous phase with no macromolecules, $\eta_c\approx10^{-3}~\mathrm{Pa~s}$), inertial effects will be negligible for droplets less than $100~\mathrm{\mu m}$ in radius.

\section{Results and Discussion}

We demonstrate sessile droplet tensiometry for three different phase separated macromolecular droplets.
The physical properties of mixtures of PEG and dextran have been thoroughly studied by other means, and therefore allow us to easily compare our approach with previous results.
We chose a far-from critical mixture of 8kDa PEG and 500kDa dextran \cite{liu}, with the dextran-rich phase making up the droplets.  
We further consider two phase-separated proteins, FUS$_{267}$ and Bik1.
FUS is a widely studied protein related to neurodegnerative diseases \cite{patel}.
It features a disordered N-terminal domain and a folded C-terminal domain.
Here, we work with just the first 267 amino acids of the disordered domain.
Finally, Bik1 is a protein found in yeast that is associated with microtubules during cell division \cite{hodek,berlin}.

\begin{figure}[ht]
\includegraphics[width = .95 \columnwidth]{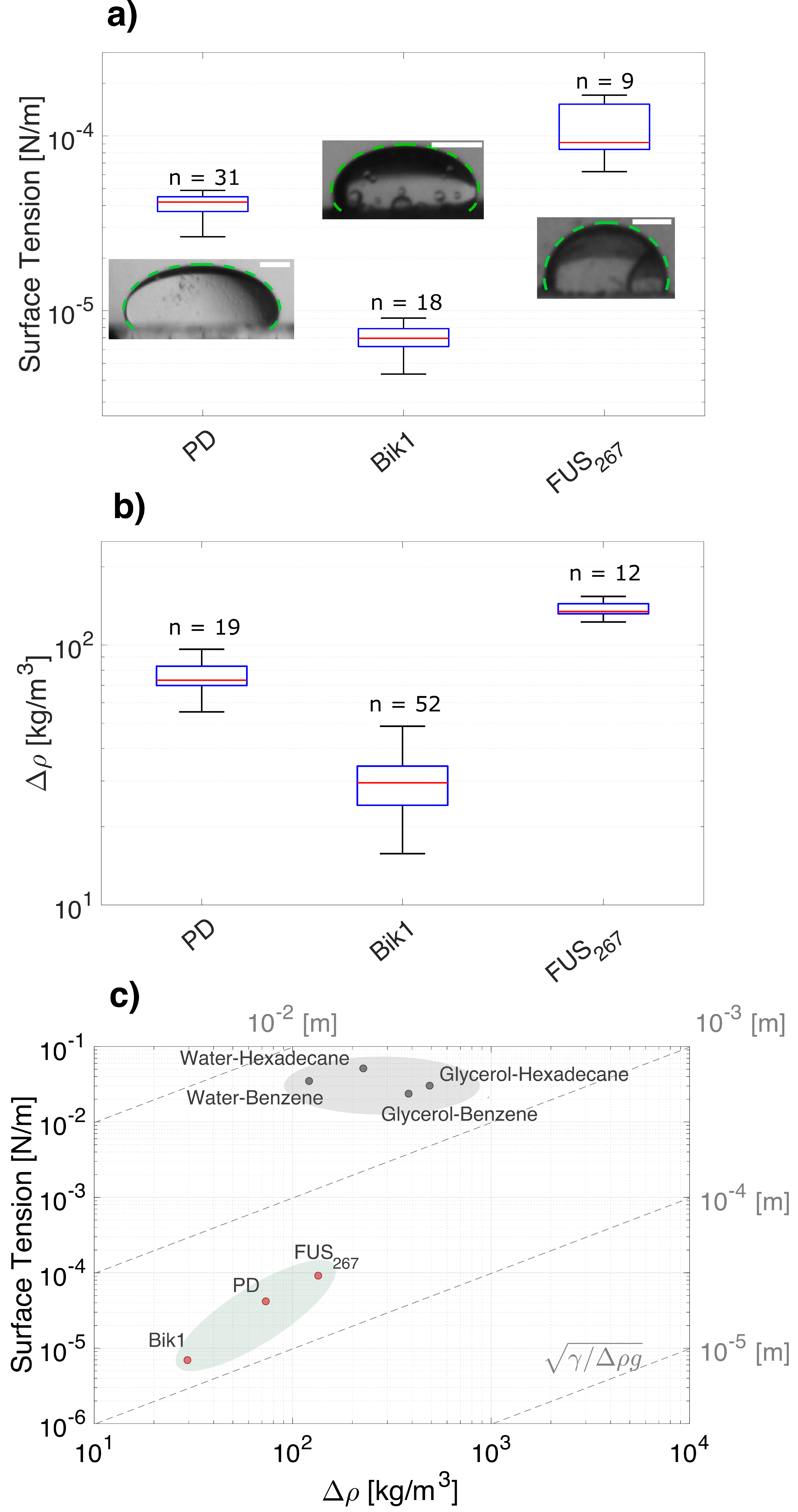}
\caption{\label{fig:cap} \emph{Surface tension, density, and the capillary length.} a) Surface tensions with example images of sessile droplets for PEG/dextran, Bik1, and FUS$_{267}$. The scale bar is 200 $\mu m$. b) density differences of the same droplets. Box plots showing median values (red lines), 25th and 75th percentiles (blue boxes), and total span of results (black bars). The number above each box plot indicates the number of droplets measured. c) Ashby plot of surface tension and density difference showing simple liquids (gray blob) and macromolecular liquids (green blob).  Dashed lines are contours of the capillary length.}
\end{figure}

\begin{figure}[t]
\centering
    \includegraphics[width = .95 \columnwidth]{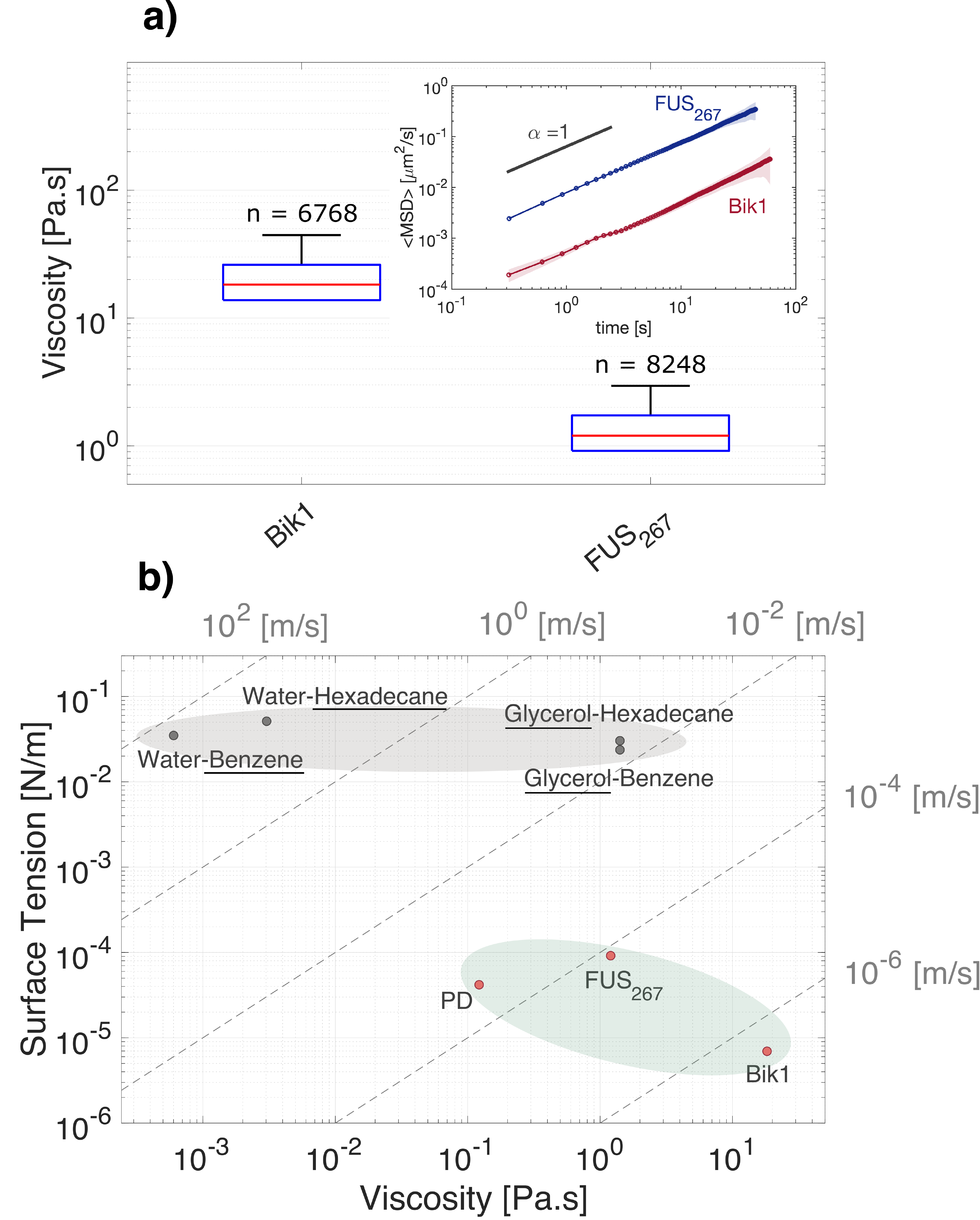}
    \caption{\label{fig:vis} \emph{Viscosity and the capillary velocity.} a) Box plot of the viscosities for Bik1 and FUS$_{267}$ with the inset mean-squared displacements of 0.1 $\mu$m radii tracers in Bik1 and FUS$_{267}$ droplets. The number above each box indicates the numbers of the tracers. b) Ashby plot of surface tension and viscosity showing simple liquids (gray blob) and macromolecular liquids (green blob). The viscosity of the underlined simple liquids are used in this graph. Dashed lines are contours of constant capillary velocity.}
\end{figure}

Measured values of the surface tension and relative density of each of these droplets is shown in Fig. \ref{fig:cap}.
The surface tensions of these systems, shown in Fig. \ref{fig:cap}a, range  from about 7 to 90 $\mathrm{\mu N/m}$.
The density differences between the two phases, shown in Fig. \ref{fig:cap}b, range from 30 to 150 $\mathrm{kg/m^3}$.
The surface tension and density values for PEG/dextran agree well with literature data \cite{liu}.
To connect sedimentation speeds of the FUS and Bik1 droplets to density differences, we assumed that the continuous phase viscosity is the same as water, and that the inner phase is much more viscous, the latter is justified by the microrheology measurements in Fig. \ref{fig:vis}.
For dextran droplets, we  measured the viscosities of both phases with a rheometer, which were 150$\pm$10 and 3.9$\pm$0.1 mPa s, for the dextran-rich and PEG-rich phases respectively. 

While the two protein systems have  similar capillary lengths, around 160 $\mu$m, they have very different surface tensions and densities.
The surface tension of Bik1 is 7 $\mathrm{\mu N / m}$, much smaller than the surface tension of  FUS$_{267}$, about 90 $\mathrm{\mu N / m}$.
There are no measurements of the surface tension of these specific protein droplets. 
The surface tension of other phase-separated protein droplets measured through other means have been reported to range  from 0.4-100 $\mu N/m$ \cite{shanacliff,clif2,Feric,Jawerth,simon}.
Bik1 droplets had a significantly smaller density mismatch than FUS$_{267}$ droplets, about $29~\mathrm{kg/m^3}$ and $135~\mathrm{kg/m^3}$, respectively.
Using a nominal protein density of  $1.3~\mathrm{g/cm^3}$, these density differences imply respective protein volume fractions within the droplets of 10$\%$ and 45$\%$.
There are few measurements of the densities of  phase-separated protein droplets.
Feric \emph{et al} \cite{Feric2} measured the density  of the nucleolus (composed of a complex mixture of proteins and nucleic acids)  using a similar sedimentation method, and found a value of 1140~$\mathrm{kg/m^3}$.
This is similar to our measured value of FUS$_{267}$ droplets, about 1135~$\mathrm{kg/m^3}$.

These values are placed in a broader context using an Ashby plot of surface tension and density in Fig. \ref{fig:cap}c.
In addition to these three droplets, we include  a selection of droplets of far-from-critical simple liquid combinations.
This plot higlights an essential feature  that contrasts macromolecular droplets (shaded light green), from simple liquids (shaded gray):  
macromolecular droplets tend to have very low surface tensions, even when they are far from critical.
Contours of the capillary length, $L_c$, are indicated by the dashed lines in Fig. \ref{fig:cap}c.
While simple liquids have capillary lengths from the millimeter to centimeter scales, our macromolecular droplets have capillary lengths between $100\mu$m and 1mm.

To complete our mechanical characterization, we  measured the rheological properties of the protein droplets using standard particle tracking microrheology  \cite{furst2017}, as described in the Materials and Methods.
Mean-squared displacements of $\approx$ 0.1 $\mu$m radius tracer particles embedded in each of the three droplet phases all show a linear dependence on time, as shown in the inset of Fig. \ref{fig:vis}.
This confirms that each droplet behaves as a simple Newtonian liquid, with no signs of viscoelasticity.
Since the mean-squared displacements are linear, the rheology of each of the droplets is simply characterized by their viscosity.
The viscosities of the Bik1 and FUS$_{267}$ droplets are 18.2 and 1.2 Pa.s, respectively.

An Ashby diagram  showing the surface tension and viscosity of droplets is shown  in Fig. \ref{fig:vis}b.
It helps to  illustrate potential advantages of the sessile droplet method over  the popular `fusion method' for determing the surface tension. 
In the latter case, the time it takes for the droplets to coalesce is set by the ratio of the droplet size and the capillary velocity $V_c=\eta_d/\gamma$.
Thus, with a knowledge of $\eta_d$, $\gamma$ can be extracted.
If the capillary velocity is too high, fusion events cannot be resolved.
The dashed lines on Fig. \ref{fig:vis}b indicate contours of constant capillary velocity.  
While the fusion of Bik1 droplets in the lower right corner of this diagram can be easily resolved, microscopic fusion experiments become much more challenging as droplets move to the top left of the diagram.
The capillary velocities of dextran and FUS$_{267}$ droplets are a hundred-fold higher, around $100~\mathrm{\mu m/s}$.
For typical $10~\mathrm{\mu m}$ droplets, resolution of fusion events would require frame rates around 100 frames per second.

\section{Conclusion}
\label{sec:conc}

Sessile droplet shape analysis is a robust method to assess the surface tension of phase separated liquid droplets, capable of measuring surface tensions over at least four orders of magnitude.
With the integration of a simple density measurement based on sedimentation velocity, this method is an efficient and effective choice for measuring the surface tensions of droplets where only microliters of material are readily available, such as condensed phases of proteins.
This simple method has the added benefit of providing easy access to the overall droplet concentration, a quantity which has been largely overlooked in the literature on phase-separated proteins.

Compared to the popular droplet fusion technique, the sessile droplet method offers several advantages.  
First, it can measure a broader range of droplets, because only the most viscous droplets will have slow enough fusion events to be resolved with standard imaging systems.  Second, it is not subject to systematic errors associated with contact line sliding \cite{hernandez2012symmetric,snoeijer2013moving}.
Third, viscoelasticity is expected to change the dynamics of droplet fusion. Sessile droplet shape analysis can work with viscoelastic droplets, as long as they have no elastic memory at long times.  

One potential weakness of the current approach is that the droplets need to partially wet the substrate.
While a simple PEG-ylated surface was sufficient for the cases studied here, different surface treatments may be needed for other droplets.
In cases where appropriate surface treatments are not readily available, a similar shape analysis of pendant droplets could be more convenient.

\section{Acknowledgements}
We acknowledge Thomas Schweizer, Solenn Riedel, and Andrea Testa for help with rheology, Tianqi Sai for help with particle sizing, Cristina Manatscha for help with cloning. We acknowledge financial support from the Swiss National Science Foundation grants 172824 (MI, A.L.T. and ERD) and 170976 (LE and FHTA) as well as the EMBO fellowship LTF-388-2018 (LE).

\section{Materials and Methods}
\subsection{Macromolecular droplets}
\par\label{PD12}\textbf{PEG/Dextran}
Polyethylene glycol (PEG) 8kDa and dextran 500kDa were purchased from Sigma Aldrich and Alfa Aeser, respectively.
We quantified the properties of dextran droplets phase separated from an aqueous mixture of  5.6wt/wt\% PEG and 7wt/wt\% Dextran \cite{liu}.
These were prepared from stock solutions of PEG (30wt/wt\% in DI water) and dextran (10wt/wt\% in DI water) that were premixed and stored at $4^{\circ} C$.

\par \textbf{FUS low complexity domain (LCD)} Histidine - and Gb1- tagged FUS (residues 1-267) was overexpressed in E.coli strain BL21 (DE3) at $20^\circ$C overnight.  The protein's molecular weight is $\approx$ 26kDa.
The protein was purified under denaturing conditions on a nickel affinity chromatography column, followed by enzymatic cleavage of the tags with TEV protease.  
Additional affinity chromatography removed the cleaved products. Finally, the protein was concentrated up to 2 mM in the presence of 6M urea and stored at $-80^\circ$C.

\par \textbf{Bik1} N-terminally tagged hexa-histidine - thrombin cleavage site - \emph{S.cerevisiae} full length Bik1 (H6-TCS-Bik1) was overexpressed in \emph{E. coli} strain BL21-CodonPlus (DE3)-RIPL (Agilent) at $20^\circ$C overnight. Protein samples were purified by nickel affinity and size exclusion chromatography in 20 mM Tris pH 7.5, 500 mM NaCl, 10\% glycerol supplemented with 10 mM Imidazole and 10 mM 2-mercaptoethanol (nickel affinity) or 1 mM DTT (size exclusion). H6-TCS-Bik1 protein samples were concentrated up to 265 $\mu$M and stored at -$80^\circ$C in size exclusion buffer.The molecular weight of this protein is $\approx$ 52kDa.

\subsection{\label{sec:passivation}Surface Passivation}

In order to characterize droplet properties it is important to create surfaces that the droplets do not spread on.
This both allows the creation of stable droplets with a finite contact angle, and avoids losing protein from the system as it adheres to any free surfaces in the sample chamber.
We create such surfaces by passivating the surface of glass coverslips with a PEG silane, (3-[Methoxy(polyethyleneoxy)propyl]trimethoxysilane, $6-9$ PEG-units) bought from ABCR GmbH.

To coat the coverslips, we immerse them using a glass staining rack in a constantly-stirred solution containing $2300mg$ of PEG silane, $500 mL$ of toluene, and $800 \mu l$ of aqueous hydrochloric acid (Hydrochloric acid fuming $37\%$ from VWR, MilliporeSigma).
After 18 hours at room temperature, the coverslips are rinsed once in toluene (toluene 99.7+$\%$ from Alfa Aesar), and then twice more with ethanol (Ethanol, denatured with IPA from Alcosuisse), before being dried with clean, dry air.
Finally coverslips are stored in a dry chamber with a dessicant until required for use.

\subsection{\label{sec:FD}Sedimenting and Sessile Droplets}
\par
\begin{figure}[t]
\includegraphics[width = .8 \columnwidth]{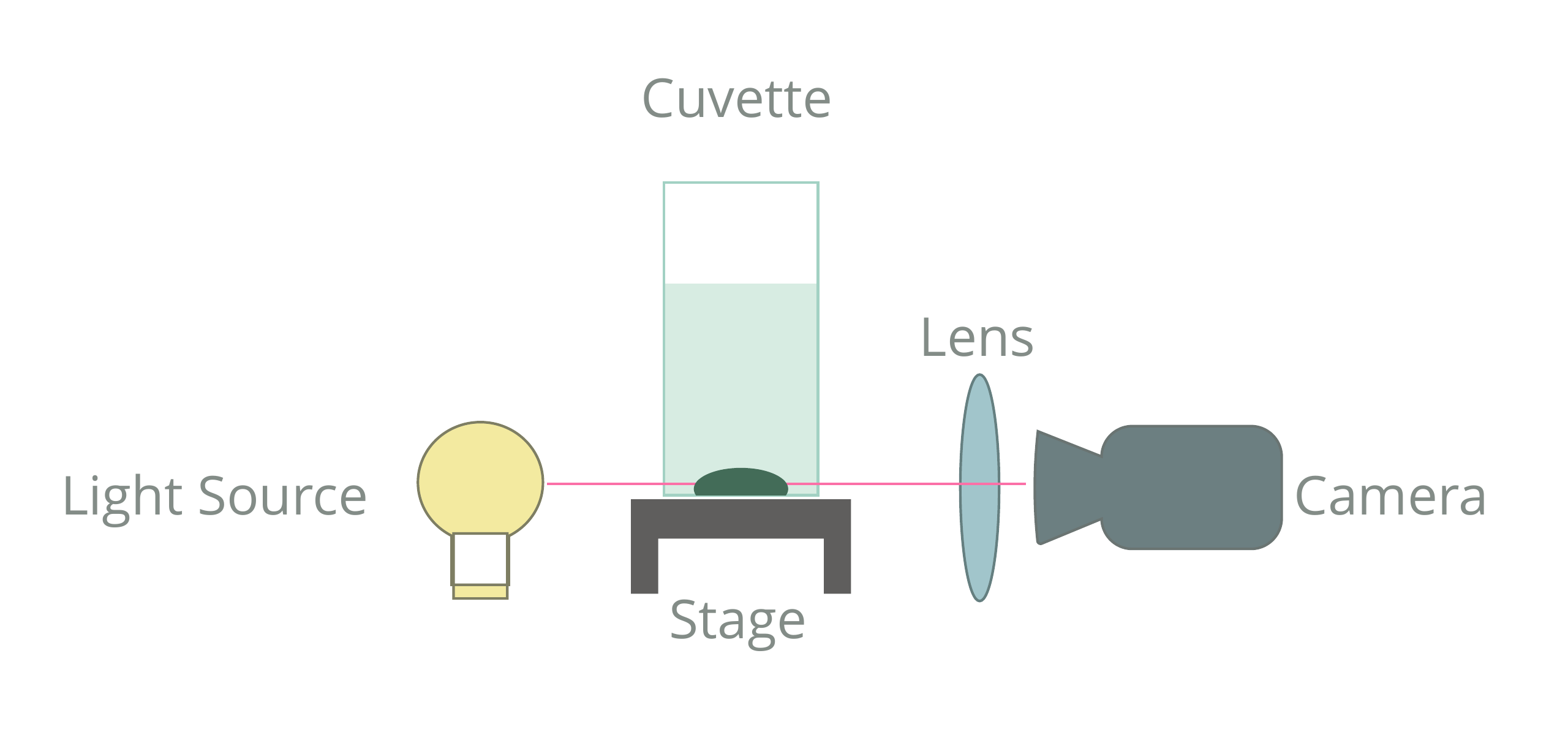}
\caption{\label{fig:angela}  A schematic of the experimental setup for falling and sessile droplets experiments. }
\end{figure}

 The optical setup for these experiments is shown in Fig. \ref{fig:angela}. 
In this figure, light source is a $3.5" \times 6"$ White, LED Backlight (Edmundoptics), the lens is a $0.5\times - 1.0\times$ VariMagTL™ Telecentric Lens (Edmundoptics) and the camera is a CMOS Camera from Thorlabs (DCC3240M - High-Sensitivity USB 3.0, 1280 x 1024, Global Shutter, Monochrome Sensor). 
Note the use of a telecentric lens, which ensures accurate size measurements. 
Samples are held in a 2.5 mL $12.5\times12.5\times45~\mathrm{mm}^3$  cuvette, with a PEG-ylated coverslip placed inside.

The cuvette is fill with about 200 $\mathrm{\mu L}$ of the continuous phase, which is separated from the droplet phase by centrifugation (10 min at 14 rcf).
Droplet phase or a dilute mixture of droplets in the continuous phase are added from the top with a micropipette.
For sedimentation experiments, a series of images is recorded as the droplets fall, typically around 3 fps.
For sessile droplet experiments, one must wait for the droplet to reach its equilibrium shape after coming in contact with the surface, which can take several minutes and depends on the size of the droplet and its capillary velocity. 

\begin{figure}[t]
\includegraphics[width = .8 \columnwidth]{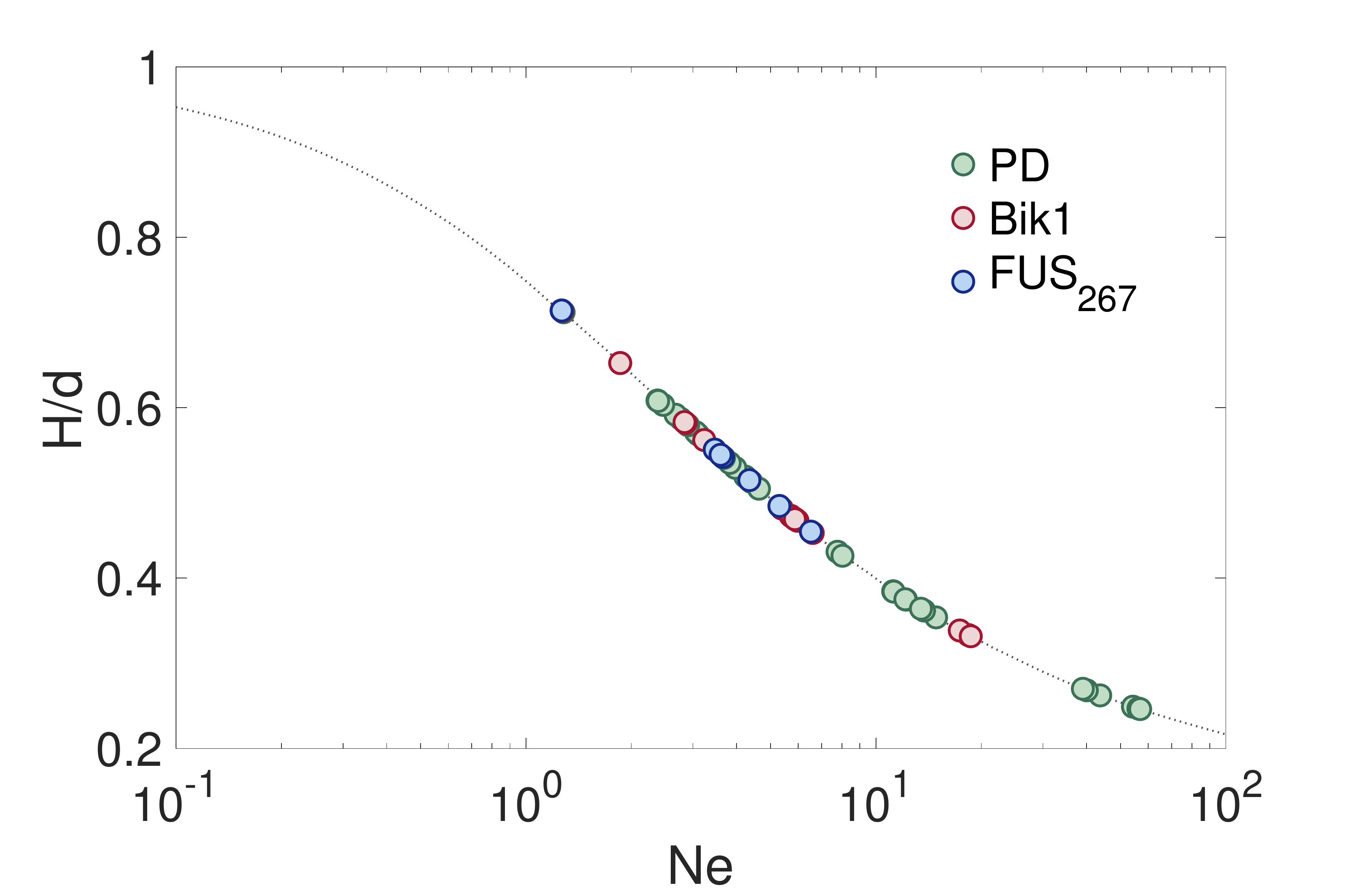}
\caption{\label{fig:neumann} The Neumann number calculated for all the sessile droplets of the 3 systems.  Note that they are all greater than 0.3.}
\end{figure}

\subsection{\label{sec:PT}Particle Tracking Microrheology}
We used $0.2\mu m$ diameter fluorescent carboxyl microspheres ($\lambda/480, 520 nm$) available at $\sim1\%$ solids ($w/v$) aqueous suspensions from Bangs Laboratories, Inc. 
To add the particles in FUS$_{267}$ and Bik1 droplets,  beads were added to the buffer used to dilute the stocks protein solutions for droplet formation. 
Beads spontaneously partitioned in the condensed (protein-rich) phase of both FUS$_{267}$ and Bik1. 
Sample chambers are constructed from PEG-ylated coverslips (section \ref{sec:passivation}) and  double sided tape, which acts both as a spacer and sealant to limit evaporation. 

Samples are imaged with fluorescence on a Nikon eclipse Ti-U inverted microscope with an Oil Nikon Apo TIRF $100\times$ objective and recorded with a ORCA Flash 4.0 v2 (digital CMOS C11440-22CU) Hamamatsu camera, controlled by $\mu$Manager.
For each protein droplet, $200$ frames we acquired with $10$ms exposures at $300$ms intervals. 

Images were analyzed in MATLAB.  Particle centers were found by fitting to a Gaussian distribution.  Particle locations were linked into trajectories using \cite{crocker}.  Diffusion coefficents were determined by fitting the mean-squared displacement versus time.
The viscosity was determined from the diffusion coefficient using the Stokes-Einstein equation.
\begin{equation}
    \label{eq:vis}
    D= k_{B}T/6\pi R\eta,
\end{equation}
where $k_{B}$ is Boltzmann constant, $T$ is the absolute temperature and $R$ is the radius of the particle.


\begin{thebibliography}{46}
\expandafter\ifx\csname natexlab\endcsname\relax\def\natexlab#1{#1}\fi
\expandafter\ifx\csname bibnamefont\endcsname\relax
  \def\bibnamefont#1{#1}\fi
\expandafter\ifx\csname bibfnamefont\endcsname\relax
  \def\bibfnamefont#1{#1}\fi
\expandafter\ifx\csname citenamefont\endcsname\relax
  \def\citenamefont#1{#1}\fi
\expandafter\ifx\csname url\endcsname\relax
  \def\url#1{\texttt{#1}}\fi
\expandafter\ifx\csname urlprefix\endcsname\relax\def\urlprefix{URL }\fi
\providecommand{\bibinfo}[2]{#2}
\providecommand{\eprint}[2][]{\url{#2}}

\bibitem[{\citenamefont{Ostwald and K{\"o}hler}(1927)}]{ostwald1927}
\bibinfo{author}{\bibfnamefont{W.}~\bibnamefont{Ostwald}} \bibnamefont{and}
  \bibinfo{author}{\bibfnamefont{R.}~\bibnamefont{K{\"o}hler}},
  \bibinfo{journal}{Kolloid-Zeitschrift} \textbf{\bibinfo{volume}{43}},
  \bibinfo{pages}{131} (\bibinfo{year}{1927}).

\bibitem[{\citenamefont{De~Jong}(1932)}]{de1932}
\bibinfo{author}{\bibfnamefont{H.~B.} \bibnamefont{De~Jong}},
  \bibinfo{journal}{Protoplasma} \textbf{\bibinfo{volume}{15}},
  \bibinfo{pages}{110} (\bibinfo{year}{1932}).

\bibitem[{\citenamefont{Overbeek and Voorn}(1957)}]{overbeek1957}
\bibinfo{author}{\bibfnamefont{J.~T.~G.} \bibnamefont{Overbeek}}
  \bibnamefont{and} \bibinfo{author}{\bibfnamefont{M.}~\bibnamefont{Voorn}},
  \bibinfo{journal}{Journal of Cellular and Comparative Physiology}
  \textbf{\bibinfo{volume}{49}}, \bibinfo{pages}{7} (\bibinfo{year}{1957}).

\bibitem[{\citenamefont{Liu et~al.}(2012)\citenamefont{Liu, Lipowsky, and
  Dimova}}]{liu}
\bibinfo{author}{\bibfnamefont{Y.}~\bibnamefont{Liu}},
  \bibinfo{author}{\bibfnamefont{R.}~\bibnamefont{Lipowsky}}, \bibnamefont{and}
  \bibinfo{author}{\bibfnamefont{R.}~\bibnamefont{Dimova}},
  \bibinfo{journal}{Langmuir} \textbf{\bibinfo{volume}{28}},
  \bibinfo{pages}{3831} (\bibinfo{year}{2012}).

\bibitem[{\citenamefont{Hyman and Brangwynne}(2011)}]{hymn1}
\bibinfo{author}{\bibfnamefont{A.~A.} \bibnamefont{Hyman}} \bibnamefont{and}
  \bibinfo{author}{\bibfnamefont{C.~P.} \bibnamefont{Brangwynne}},
  \bibinfo{journal}{Cell} \textbf{\bibinfo{volume}{21}}, \bibinfo{pages}{14}
  (\bibinfo{year}{2011}).

\bibitem[{\citenamefont{Weber and Brangwynne}(2012)}]{web}
\bibinfo{author}{\bibfnamefont{S.~C.} \bibnamefont{Weber}} \bibnamefont{and}
  \bibinfo{author}{\bibfnamefont{C.~P.} \bibnamefont{Brangwynne}},
  \bibinfo{journal}{Cell} \textbf{\bibinfo{volume}{146}}, \bibinfo{pages}{1188}
  (\bibinfo{year}{2012}).

\bibitem[{\citenamefont{Brangwynne et~al.}(2009)\citenamefont{Brangwynne,
  Eckmann, Courson, Rybarska, Hoege, Gharakhani, J{\"u}licher, and
  Hyman}}]{clif2}
\bibinfo{author}{\bibfnamefont{C.~P.} \bibnamefont{Brangwynne}},
  \bibinfo{author}{\bibfnamefont{C.~R.} \bibnamefont{Eckmann}},
  \bibinfo{author}{\bibfnamefont{D.~S.} \bibnamefont{Courson}},
  \bibinfo{author}{\bibfnamefont{A.}~\bibnamefont{Rybarska}},
  \bibinfo{author}{\bibfnamefont{C.}~\bibnamefont{Hoege}},
  \bibinfo{author}{\bibfnamefont{J.}~\bibnamefont{Gharakhani}},
  \bibinfo{author}{\bibfnamefont{F.}~\bibnamefont{J{\"u}licher}},
  \bibnamefont{and} \bibinfo{author}{\bibfnamefont{A.~A.} \bibnamefont{Hyman}},
  \bibinfo{journal}{Science} \textbf{\bibinfo{volume}{324(5935)}},
  \bibinfo{pages}{1729} (\bibinfo{year}{2009}).

\bibitem[{\citenamefont{Kedersha et~al.}(2005)\citenamefont{Kedersha,
  Stoecklin, Ayodele, Yacono, Lykke-Andersen, Fritzler, Scheuner, Kaufman,
  Golan, and Anderson}}]{ked1}
\bibinfo{author}{\bibfnamefont{N.}~\bibnamefont{Kedersha}},
  \bibinfo{author}{\bibfnamefont{G.}~\bibnamefont{Stoecklin}},
  \bibinfo{author}{\bibfnamefont{M.}~\bibnamefont{Ayodele}},
  \bibinfo{author}{\bibfnamefont{P.}~\bibnamefont{Yacono}},
  \bibinfo{author}{\bibfnamefont{J.}~\bibnamefont{Lykke-Andersen}},
  \bibinfo{author}{\bibfnamefont{M.~J.} \bibnamefont{Fritzler}},
  \bibinfo{author}{\bibfnamefont{D.}~\bibnamefont{Scheuner}},
  \bibinfo{author}{\bibfnamefont{R.~J.} \bibnamefont{Kaufman}},
  \bibinfo{author}{\bibfnamefont{D.~E.} \bibnamefont{Golan}}, \bibnamefont{and}
  \bibinfo{author}{\bibfnamefont{P.}~\bibnamefont{Anderson}},
  \bibinfo{journal}{The Journal of Cell Biology}
  \textbf{\bibinfo{volume}{169(6)}}, \bibinfo{pages}{871}
  (\bibinfo{year}{2005}).

\bibitem[{\citenamefont{Strzelecka et~al.}(2010)\citenamefont{Strzelecka,
  Trowitzch, Weber, Lurmann, Oates, and Neugebauer}}]{stz}
\bibinfo{author}{\bibfnamefont{M.}~\bibnamefont{Strzelecka}},
  \bibinfo{author}{\bibfnamefont{S.}~\bibnamefont{Trowitzch}},
  \bibinfo{author}{\bibfnamefont{G.}~\bibnamefont{Weber}},
  \bibinfo{author}{\bibfnamefont{R.}~\bibnamefont{Lurmann}},
  \bibinfo{author}{\bibfnamefont{A.~C.} \bibnamefont{Oates}}, \bibnamefont{and}
  \bibinfo{author}{\bibfnamefont{K.~M.} \bibnamefont{Neugebauer}},
  \bibinfo{journal}{Nature Structural and Molecular Biology}
  \textbf{\bibinfo{volume}{17}}, \bibinfo{pages}{403?409}
  (\bibinfo{year}{2010}).

\bibitem[{\citenamefont{Brangwynne et~al.}(2010)\citenamefont{Brangwynne,
  Mitchison, and Hyman}}]{clif3}
\bibinfo{author}{\bibfnamefont{C.~P.} \bibnamefont{Brangwynne}},
  \bibinfo{author}{\bibfnamefont{T.~J.} \bibnamefont{Mitchison}},
  \bibnamefont{and} \bibinfo{author}{\bibfnamefont{A.~A.} \bibnamefont{Hyman}},
  \bibinfo{journal}{PNAS} \textbf{\bibinfo{volume}{108(11)}},
  \bibinfo{pages}{4334} (\bibinfo{year}{2010}).

\bibitem[{\citenamefont{Wright and Dyson}(2015)}]{wright}
\bibinfo{author}{\bibfnamefont{P.~E.} \bibnamefont{Wright}} \bibnamefont{and}
  \bibinfo{author}{\bibfnamefont{H.~J.} \bibnamefont{Dyson}},
  \bibinfo{journal}{Nat. Rev. Mol. Cell Biol.}
  \textbf{\bibinfo{volume}{16(1)}}, \bibinfo{pages}{18} (\bibinfo{year}{2015}).

\bibitem[{\citenamefont{Dunker et~al.}(2005)\citenamefont{Dunker, Cortese,
  Romero, Iakoucheva, and Uversky}}]{dunker}
\bibinfo{author}{\bibfnamefont{A.~K.} \bibnamefont{Dunker}},
  \bibinfo{author}{\bibfnamefont{M.~S.} \bibnamefont{Cortese}},
  \bibinfo{author}{\bibfnamefont{P.}~\bibnamefont{Romero}},
  \bibinfo{author}{\bibfnamefont{L.~M.} \bibnamefont{Iakoucheva}},
  \bibnamefont{and} \bibinfo{author}{\bibfnamefont{V.~N.}
  \bibnamefont{Uversky}}, \bibinfo{journal}{FEBS}
  \textbf{\bibinfo{volume}{272(20)}}, \bibinfo{pages}{5129}
  (\bibinfo{year}{2005}).

\bibitem[{\citenamefont{Dyson and Wright}(2002)}]{dyson}
\bibinfo{author}{\bibfnamefont{J.~H.} \bibnamefont{Dyson}} \bibnamefont{and}
  \bibinfo{author}{\bibfnamefont{P.~E.} \bibnamefont{Wright}},
  \bibinfo{journal}{Curr. Opin. Struct. Biol.}
  \textbf{\bibinfo{volume}{12(1)}}, \bibinfo{pages}{54} (\bibinfo{year}{2002}).

\bibitem[{\citenamefont{Tompa and Fersht}(2009)}]{tomp}
\bibinfo{author}{\bibfnamefont{P.}~\bibnamefont{Tompa}} \bibnamefont{and}
  \bibinfo{author}{\bibfnamefont{A.}~\bibnamefont{Fersht}},
  \bibinfo{journal}{Chapman and Hall/CRC}  (\bibinfo{year}{2009}).

\bibitem[{\citenamefont{Vekilov}(2010)}]{vek}
\bibinfo{author}{\bibfnamefont{P.~G.} \bibnamefont{Vekilov}},
  \bibinfo{journal}{Soft Matter} \textbf{\bibinfo{volume}{6}},
  \bibinfo{pages}{5254} (\bibinfo{year}{2010}).

\bibitem[{\citenamefont{Weber et~al.}(2014)\citenamefont{Weber, Jack,
  Schwantes, and Pande}}]{web2}
\bibinfo{author}{\bibfnamefont{J.~K.} \bibnamefont{Weber}},
  \bibinfo{author}{\bibfnamefont{R.~L.} \bibnamefont{Jack}},
  \bibinfo{author}{\bibfnamefont{C.~R.} \bibnamefont{Schwantes}},
  \bibnamefont{and} \bibinfo{author}{\bibfnamefont{V.~S.} \bibnamefont{Pande}},
  \bibinfo{journal}{Biophysical Journal} \textbf{\bibinfo{volume}{107(4)}},
  \bibinfo{pages}{974} (\bibinfo{year}{2014}).

\bibitem[{\citenamefont{Kedersha et~al.}(2013)\citenamefont{Kedersha, Ivanov,
  and Anderson}}]{ked2}
\bibinfo{author}{\bibfnamefont{N.}~\bibnamefont{Kedersha}},
  \bibinfo{author}{\bibfnamefont{P.}~\bibnamefont{Ivanov}}, \bibnamefont{and}
  \bibinfo{author}{\bibfnamefont{P.}~\bibnamefont{Anderson}},
  \bibinfo{journal}{Trends in Biochemical Sciences}
  \textbf{\bibinfo{volume}{36(10)}}, \bibinfo{pages}{494?506}
  (\bibinfo{year}{2013}).

\bibitem[{\citenamefont{Tompa and Csermely}(2004)}]{tomp2}
\bibinfo{author}{\bibfnamefont{P.}~\bibnamefont{Tompa}} \bibnamefont{and}
  \bibinfo{author}{\bibfnamefont{P.}~\bibnamefont{Csermely}},
  \bibinfo{journal}{FASEB} \textbf{\bibinfo{volume}{18(11)}},
  \bibinfo{pages}{1169?1175} (\bibinfo{year}{2004}).

\bibitem[{\citenamefont{Furst and Squires}(2017)}]{furst2017}
\bibinfo{author}{\bibfnamefont{E.~M.} \bibnamefont{Furst}} \bibnamefont{and}
  \bibinfo{author}{\bibfnamefont{T.~M.} \bibnamefont{Squires}},
  \emph{\bibinfo{title}{Microrheology}} (\bibinfo{publisher}{Oxford University
  Press}, \bibinfo{year}{2017}).

\bibitem[{\citenamefont{Taylor et~al.}(2016)\citenamefont{Taylor,
  Elbaum-Garfinkle, Vaidya, Zhang, Stoneb, and Brangwynne}}]{Taylor}
\bibinfo{author}{\bibfnamefont{N.}~\bibnamefont{Taylor}},
  \bibinfo{author}{\bibfnamefont{S.}~\bibnamefont{Elbaum-Garfinkle}},
  \bibinfo{author}{\bibfnamefont{N.}~\bibnamefont{Vaidya}},
  \bibinfo{author}{\bibfnamefont{H.}~\bibnamefont{Zhang}},
  \bibinfo{author}{\bibfnamefont{H.~A.} \bibnamefont{Stoneb}},
  \bibnamefont{and} \bibinfo{author}{\bibfnamefont{C.~P.}
  \bibnamefont{Brangwynne}}, \bibinfo{journal}{Soft Matter}
  \textbf{\bibinfo{volume}{12}}, \bibinfo{pages}{9142} (\bibinfo{year}{2016}).

\bibitem[{\citenamefont{Murakami et~al.}(2015)\citenamefont{Murakami, Qamar,
  and et~al}}]{Murakami}
\bibinfo{author}{\bibfnamefont{T.}~\bibnamefont{Murakami}},
  \bibinfo{author}{\bibfnamefont{S.}~\bibnamefont{Qamar}}, \bibnamefont{and}
  \bibinfo{author}{\bibfnamefont{J.~Q.~L.} \bibnamefont{et~al}},
  \bibinfo{journal}{Neuron} \textbf{\bibinfo{volume}{88(4)}},
  \bibinfo{pages}{678} (\bibinfo{year}{2015}).

\bibitem[{\citenamefont{Elbaum-Garfinkle
  et~al.}(2015)\citenamefont{Elbaum-Garfinkle, Younghoon~Kim, and
  et~al.}}]{shanacliff}
\bibinfo{author}{\bibfnamefont{S.}~\bibnamefont{Elbaum-Garfinkle}},
  \bibinfo{author}{\bibfnamefont{K.~S.} \bibnamefont{Younghoon~Kim}},
  \bibnamefont{and} \bibinfo{author}{\bibnamefont{et~al.}},
  \bibinfo{journal}{PNAS} \textbf{\bibinfo{volume}{112}},
  \bibinfo{pages}{7189?7194} (\bibinfo{year}{2015}).

\bibitem[{\citenamefont{Hern{\'a}ndez-S{\'a}nchez
  et~al.}(2012)\citenamefont{Hern{\'a}ndez-S{\'a}nchez, Lubbers, Eddi, and
  Snoeijer}}]{hernandez2012symmetric}
\bibinfo{author}{\bibfnamefont{J.}~\bibnamefont{Hern{\'a}ndez-S{\'a}nchez}},
  \bibinfo{author}{\bibfnamefont{L.}~\bibnamefont{Lubbers}},
  \bibinfo{author}{\bibfnamefont{A.}~\bibnamefont{Eddi}}, \bibnamefont{and}
  \bibinfo{author}{\bibfnamefont{J.}~\bibnamefont{Snoeijer}},
  \bibinfo{journal}{Phys. Rev. Lett.} \textbf{\bibinfo{volume}{109}},
  \bibinfo{pages}{184502} (\bibinfo{year}{2012}).

\bibitem[{\citenamefont{Blázquez-Castro}(2019)}]{alfonso}
\bibinfo{author}{\bibfnamefont{A.}~\bibnamefont{Blázquez-Castro}},
  \bibinfo{journal}{Micromachine} \textbf{\bibinfo{volume}{10}},
  \bibinfo{pages}{507} (\bibinfo{year}{2019}).

\bibitem[{\citenamefont{Jawerth et~al.}(2018)\citenamefont{Jawerth, Ijavi,
  Ruer, Saha, Jahnel, Hyman, Jülicher, and Fischer-Friedrich}}]{Jawerth}
\bibinfo{author}{\bibfnamefont{L.~M.} \bibnamefont{Jawerth}},
  \bibinfo{author}{\bibfnamefont{M.}~\bibnamefont{Ijavi}},
  \bibinfo{author}{\bibfnamefont{M.}~\bibnamefont{Ruer}},
  \bibinfo{author}{\bibfnamefont{S.}~\bibnamefont{Saha}},
  \bibinfo{author}{\bibfnamefont{M.}~\bibnamefont{Jahnel}},
  \bibinfo{author}{\bibfnamefont{A.~A.} \bibnamefont{Hyman}},
  \bibinfo{author}{\bibfnamefont{F.}~\bibnamefont{Jülicher}},
  \bibnamefont{and}
  \bibinfo{author}{\bibfnamefont{E.}~\bibnamefont{Fischer-Friedrich}},
  \bibinfo{journal}{Phys. Rev. Lett.} \textbf{\bibinfo{volume}{121}},
  \bibinfo{pages}{258101} (\bibinfo{year}{2018}).

\bibitem[{\citenamefont{Zitzler et~al.}(2002)\citenamefont{Zitzler,
  Herminghaus, and Mugele}}]{zitzler2002capillary}
\bibinfo{author}{\bibfnamefont{L.}~\bibnamefont{Zitzler}},
  \bibinfo{author}{\bibfnamefont{S.}~\bibnamefont{Herminghaus}},
  \bibnamefont{and} \bibinfo{author}{\bibfnamefont{F.}~\bibnamefont{Mugele}},
  \bibinfo{journal}{Physical Review B} \textbf{\bibinfo{volume}{66}},
  \bibinfo{pages}{155436} (\bibinfo{year}{2002}).

\bibitem[{\citenamefont{Anachkov et~al.}(2016)\citenamefont{Anachkov, Lesov,
  Zanini, Kralchevsky, Denkov, and Isa}}]{anachkov2016}
\bibinfo{author}{\bibfnamefont{S.~E.} \bibnamefont{Anachkov}},
  \bibinfo{author}{\bibfnamefont{I.}~\bibnamefont{Lesov}},
  \bibinfo{author}{\bibfnamefont{M.}~\bibnamefont{Zanini}},
  \bibinfo{author}{\bibfnamefont{P.~A.} \bibnamefont{Kralchevsky}},
  \bibinfo{author}{\bibfnamefont{N.~D.} \bibnamefont{Denkov}},
  \bibnamefont{and} \bibinfo{author}{\bibfnamefont{L.}~\bibnamefont{Isa}},
  \bibinfo{journal}{Soft Matter} \textbf{\bibinfo{volume}{12}},
  \bibinfo{pages}{7632} (\bibinfo{year}{2016}).

\bibitem[{\citenamefont{Saad and Neumann}(2016)}]{saad2016}
\bibinfo{author}{\bibfnamefont{S.~M.} \bibnamefont{Saad}} \bibnamefont{and}
  \bibinfo{author}{\bibfnamefont{A.~W.} \bibnamefont{Neumann}},
  \bibinfo{journal}{Adv. Colloid Interface Sci.}
  \textbf{\bibinfo{volume}{238}}, \bibinfo{pages}{62} (\bibinfo{year}{2016}).

\bibitem[{\citenamefont{Atefi et~al.}(2014)\citenamefont{Atefi, Jr, and
  Tavana}}]{Atefi}
\bibinfo{author}{\bibfnamefont{E.}~\bibnamefont{Atefi}},
  \bibinfo{author}{\bibfnamefont{J.~A.~M.} \bibnamefont{Jr}}, \bibnamefont{and}
  \bibinfo{author}{\bibfnamefont{H.}~\bibnamefont{Tavana}},
  \bibinfo{journal}{Langmuir} \textbf{\bibinfo{volume}{30}},
  \bibinfo{pages}{9691} (\bibinfo{year}{2014}).

\bibitem[{\citenamefont{Feric et~al.}(2016)\citenamefont{Feric, Vaidya, and
  et~al.}}]{Feric}
\bibinfo{author}{\bibfnamefont{M.}~\bibnamefont{Feric}},
  \bibinfo{author}{\bibfnamefont{N.}~\bibnamefont{Vaidya}}, \bibnamefont{and}
  \bibinfo{author}{\bibfnamefont{T.~S.~H.} \bibnamefont{et~al.}},
  \bibinfo{journal}{Cell} \textbf{\bibinfo{volume}{165}}, \bibinfo{pages}{1686}
  (\bibinfo{year}{2016}).

\bibitem[{\citenamefont{Feric and Brangwynne}(2013)}]{Feric2}
\bibinfo{author}{\bibfnamefont{M.}~\bibnamefont{Feric}} \bibnamefont{and}
  \bibinfo{author}{\bibfnamefont{C.~P.} \bibnamefont{Brangwynne}},
  \bibinfo{journal}{Nature Cell Biology} \textbf{\bibinfo{volume}{15}},
  \bibinfo{pages}{1253} (\bibinfo{year}{2013}).

\bibitem[{\citenamefont{Sefiane et~al.}(2003)\citenamefont{Sefiane, Tadrist,
  and Douglas}}]{sefiane}
\bibinfo{author}{\bibfnamefont{K.}~\bibnamefont{Sefiane}},
  \bibinfo{author}{\bibfnamefont{L.}~\bibnamefont{Tadrist}}, \bibnamefont{and}
  \bibinfo{author}{\bibfnamefont{M.}~\bibnamefont{Douglas}},
  \bibinfo{journal}{International Journal of Heat and Mass Transfer}
  \textbf{\bibinfo{volume}{46}}, \bibinfo{pages}{4527?4534}
  (\bibinfo{year}{2003}).

\bibitem[{\citenamefont{Drelich}(2013)}]{drelich}
\bibinfo{author}{\bibfnamefont{J.}~\bibnamefont{Drelich}},
  \bibinfo{journal}{Surface Innovations} \textbf{\bibinfo{volume}{1}},
  \bibinfo{pages}{248?254} (\bibinfo{year}{2013}).

\bibitem[{\citenamefont{Del~R{\i}o and Neumann}(1997)}]{del1997axisymmetric}
\bibinfo{author}{\bibfnamefont{O.}~\bibnamefont{Del~R{\i}o}} \bibnamefont{and}
  \bibinfo{author}{\bibfnamefont{A.}~\bibnamefont{Neumann}},
  \bibinfo{journal}{J. Colloid Interface Sci.} \textbf{\bibinfo{volume}{196}},
  \bibinfo{pages}{136} (\bibinfo{year}{1997}).

\bibitem[{\citenamefont{Yang et~al.}(2017)\citenamefont{Yang, Yu, and
  Zuo}}]{yang2017accuracy}
\bibinfo{author}{\bibfnamefont{J.}~\bibnamefont{Yang}},
  \bibinfo{author}{\bibfnamefont{K.}~\bibnamefont{Yu}}, \bibnamefont{and}
  \bibinfo{author}{\bibfnamefont{Y.~Y.} \bibnamefont{Zuo}},
  \bibinfo{journal}{Langmuir} \textbf{\bibinfo{volume}{33}},
  \bibinfo{pages}{8914} (\bibinfo{year}{2017}).

\bibitem[{\citenamefont{Hadamard}(1911)}]{hadamard1911mouvement}
\bibinfo{author}{\bibfnamefont{J.}~\bibnamefont{Hadamard}},
  \bibinfo{journal}{C. R. Acad. Sci.} \textbf{\bibinfo{volume}{152}},
  \bibinfo{pages}{1735} (\bibinfo{year}{1911}).

\bibitem[{\citenamefont{Rallison}(1984)}]{rallison1984deformation}
\bibinfo{author}{\bibfnamefont{J.}~\bibnamefont{Rallison}},
  \bibinfo{journal}{Ann. Rev. Fluid Mechanics} \textbf{\bibinfo{volume}{16}},
  \bibinfo{pages}{45} (\bibinfo{year}{1984}).

\bibitem[{\citenamefont{Stone}(1994)}]{stone1994}
\bibinfo{author}{\bibfnamefont{H.~A.} \bibnamefont{Stone}},
  \bibinfo{journal}{Annual Review of Fluid Mechanics}
  \textbf{\bibinfo{volume}{26}}, \bibinfo{pages}{65} (\bibinfo{year}{1994}).

\bibitem[{\citenamefont{Davis and Acrivos}(1985)}]{davis1985sedimentation}
\bibinfo{author}{\bibfnamefont{R.~H.} \bibnamefont{Davis}} \bibnamefont{and}
  \bibinfo{author}{\bibfnamefont{A.}~\bibnamefont{Acrivos}},
  \bibinfo{journal}{Annual Review of Fluid Mechanics}
  \textbf{\bibinfo{volume}{17}}, \bibinfo{pages}{91} (\bibinfo{year}{1985}).

\bibitem[{\citenamefont{Segre et~al.}(1997)\citenamefont{Segre, Herbolzheimer,
  and Chaikin}}]{segre1997long}
\bibinfo{author}{\bibfnamefont{P.}~\bibnamefont{Segre}},
  \bibinfo{author}{\bibfnamefont{E.}~\bibnamefont{Herbolzheimer}},
  \bibnamefont{and} \bibinfo{author}{\bibfnamefont{P.}~\bibnamefont{Chaikin}},
  \bibinfo{journal}{Physical Review Letters} \textbf{\bibinfo{volume}{79}},
  \bibinfo{pages}{2574} (\bibinfo{year}{1997}).

\bibitem[{\citenamefont{Patel et~al.}(2015)\citenamefont{Patel, Lee, Jawerth,
  and et~al.}}]{patel}
\bibinfo{author}{\bibfnamefont{A.}~\bibnamefont{Patel}},
  \bibinfo{author}{\bibfnamefont{H.~O.} \bibnamefont{Lee}},
  \bibinfo{author}{\bibfnamefont{L.}~\bibnamefont{Jawerth}}, \bibnamefont{and}
  \bibinfo{author}{\bibnamefont{et~al.}}, \bibinfo{journal}{Cell}
  \textbf{\bibinfo{volume}{162}}, \bibinfo{pages}{1066} (\bibinfo{year}{2015}).

\bibitem[{\citenamefont{Blake-Hodek et~al.}(2010)\citenamefont{Blake-Hodek,
  Cassimeris, and Huffaker}}]{hodek}
\bibinfo{author}{\bibfnamefont{K.~A.} \bibnamefont{Blake-Hodek}},
  \bibinfo{author}{\bibfnamefont{L.}~\bibnamefont{Cassimeris}},
  \bibnamefont{and} \bibinfo{author}{\bibfnamefont{T.~C.}
  \bibnamefont{Huffaker}}, \bibinfo{journal}{Molecular Biology of the Cell}
  \textbf{\bibinfo{volume}{21}}, \bibinfo{pages}{2013?2023}
  (\bibinfo{year}{2010}).

\bibitem[{\citenamefont{Berlin et~al.}(1990)\citenamefont{Berlin, Styles, and
  Fink}}]{berlin}
\bibinfo{author}{\bibfnamefont{V.}~\bibnamefont{Berlin}},
  \bibinfo{author}{\bibfnamefont{C.~A.} \bibnamefont{Styles}},
  \bibnamefont{and} \bibinfo{author}{\bibfnamefont{G.~R.} \bibnamefont{Fink}},
  \bibinfo{journal}{Cell Biology} \textbf{\bibinfo{volume}{111}},
  \bibinfo{pages}{2573} (\bibinfo{year}{1990}).

\bibitem[{\citenamefont{Alberti et~al.}(2019)\citenamefont{Alberti, Gladfelter,
  and Mittag}}]{simon}
\bibinfo{author}{\bibfnamefont{S.}~\bibnamefont{Alberti}},
  \bibinfo{author}{\bibfnamefont{A.}~\bibnamefont{Gladfelter}},
  \bibnamefont{and} \bibinfo{author}{\bibfnamefont{T.}~\bibnamefont{Mittag}},
  \bibinfo{journal}{Cell} \textbf{\bibinfo{volume}{176}}, \bibinfo{pages}{419}
  (\bibinfo{year}{2019}).

\bibitem[{\citenamefont{Snoeijer and Andreotti}(2013)}]{snoeijer2013moving}
\bibinfo{author}{\bibfnamefont{J.~H.} \bibnamefont{Snoeijer}} \bibnamefont{and}
  \bibinfo{author}{\bibfnamefont{B.}~\bibnamefont{Andreotti}},
  \bibinfo{journal}{Annual review of fluid mechanics}
  \textbf{\bibinfo{volume}{45}} (\bibinfo{year}{2013}).

\bibitem[{\citenamefont{Crocker and Grier}(1996)}]{crocker}
\bibinfo{author}{\bibfnamefont{J.~C.} \bibnamefont{Crocker}} \bibnamefont{and}
  \bibinfo{author}{\bibfnamefont{D.~G.} \bibnamefont{Grier}},
  \bibinfo{journal}{Journal of Colloid and Interface Science}
  \textbf{\bibinfo{volume}{179}}, \bibinfo{pages}{298} (\bibinfo{year}{1996}).

\end{thebibliography}

\providecommand{\noopsort}[1]{}\providecommand{\singleletter}[1]{#1}%

\end{document}